\documentclass[aps,prl,twocolumn,superscriptaddress,floatfix,citeautoscript]{revtex4}
\usepackage{times}
\usepackage{graphicx}
\usepackage{dcolumn}
\usepackage{bm}
\usepackage{color}
\usepackage{lipsum}
\usepackage{amssymb}   
\newcommand {\beq} {\begin{equation}}
\newcommand {\eeq} {\end{equation}}
\newcommand {\bqa} {\begin{eqnarray}}
\newcommand {\eqa} {\end{eqnarray}}

\usepackage{hyperref}
\begin{document}

\title{New paradigm for a disordered superconductor in a magnetic field}

\author{Anushree Datta}
\affiliation{Indian Institute of Science Education and Research Kolkata, Mohanpur, India-741246}

\author{Anurag Banerjee\footnote{Present address: Institut de Physique Th\'eorique, Universit\'e Paris-Saclay, CEA, CNRS, F-91191 Gif-sur-Yvette, France.}}
\affiliation{Indian Institute of Science Education and Research Kolkata, Mohanpur, India-741246}

\author{Nandini Trivedi}
\affiliation{Department of Physics, The Ohio State University, Columbus, Ohio 43210, USA}

\author{Amit Ghosal}
\affiliation{Indian Institute of Science Education and Research Kolkata, Mohanpur, India-741246}

\begin{abstract}

We show that while orbital magnetic field and disorder, acting individually weaken superconductivity, acting together they produce an intriguing evolution of a two-dimensional type-II s-wave superconductor. 
For weak disorder, the critical field $H_c$ at which the superfluid density collapses is coincident with the field at which the superconducting energy gap gets suppressed. However, with increasing disorder these two fields diverge from each other creating a pseudogap region. The nature of vortices also transform from Abrikosov vortices with a metallic core for weak disorder to Josephson vortices with gapped and insulating cores for higher disorder. Our results naturally explain two outstanding puzzles: (1) the gigantic magnetoresistance peak observed as a function of magnetic field in thin disordered superconducting films; and (2) the disappearance of the celebrated zero-bias Caroli-de Gennes-Matricon (CdGM) peak in disordered superconductors.

\end{abstract}

\maketitle

\noindent {\it Introduction:} --
The response of an s-wave superconductor (sSC) individually to disorder and orbital magnetic field has by now been well established~\cite{Book1,Book2,RevModPhys.66.1125,RevModPhys.76.975}. In a pristine Bardeen-Cooper-Schrieffer (BCS) superconductor, the two energy scales, the single particle energy gap, $E_g$ measurable by scanning tunneling spectroscopy~\cite{RMP79.353}, and the superfluid stiffness, $D_s$, related to the diamagnetic susceptibility~\cite{Tinkham}, both vanish simultaneously at the critical temperature $T_c$.  

\begin{figure}[h]
{\includegraphics[width=8.5cm,keepaspectratio]{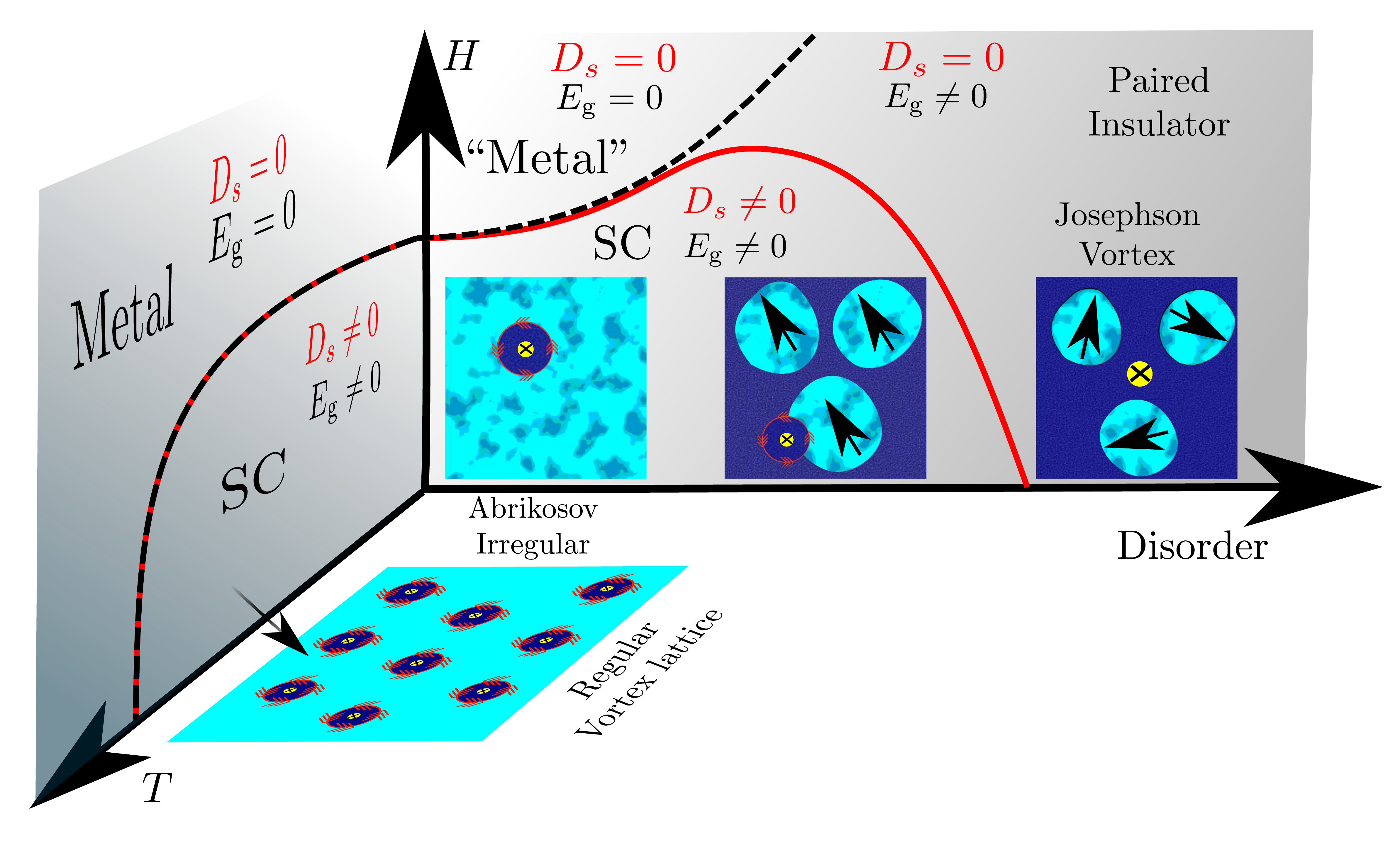}}
\caption{(Color online) {\bf Schematic Phase diagram} of a type II superconductor (SC) in the magnetic field (H), temperature (T), and disorder (V) planes. In the HT plane, the clean SC is in a Abrikosov vortex state ($H_{c1}=0$). Vortices with a metallic core form a  triangular lattice, as shown. The vortex phase transitions to a metal either by increasing $H$ or $T$. Both the superfluid stiffness $D_s$ (red) and the energy gap $E_g$ (black) vanish simultaneously on the black-red transition lines. 
Our present study unveils the $HV$-plane: For weak disorder, there is a SC to metal transition where $D_s$ and $E_g$ vanish simultaneously. A new paradigm emerges for intermediate to high disorder with the following features: (a) The critical field at which $E_g$ vanishes (black dashed line) starts diverging from the critical field at which $D_s$ (red line) vanishes. (b) The vortices change character from Abrikosov-type with metallic cores to Josephson- type with insulating cores. 
These features naturally explain puzzling experimental signatures in magneto-transport and spectroscopy. 
}
\label{fig:fig1}
\end{figure}

Upon including disorder, extensive research in last few decades has established that the pairing amplitude of a disordered superconductor becomes inhomogeneous forming SC-islands on the scale of the coherence length, $\xi$, separated by an insulating sea~\cite{PRL81.3940, PhysRevB.65.014501, NatPhys7.884}.
Ultimately, the superconductor is driven into an insulating state not by the collapse of the single particle energy gap~\cite{Stewart1273, PhysRevB.75.184530, Sacepe2011,PhysRevLett.108.177006} but rather by the collapse of the superfluid phase stiffness (See Fig.~\ref{fig:fig1}(a)) due to enhanced quantum
phase fluctuations~\cite{PRB71.014514,Nature449.876,PhysRevLett.106.047001}. Since $E_g$ remains finite while $D_s$ collapses, it is argued that the universal properties near the superconductor-insulator transition (SIT) is well described by an effective ``bosonic" Hamiltonian~\cite{PhysRevLett.64.587,PhysRevB.44.6883,PhysRevX.4.021007}.

Turning next to the effect of an applied magnetic field, $H$, on a clean BCS superconductor, it is well known that the magnetic field penetrates the SC by generating a periodic array of Abrikosov vortices~\cite{AVL} with a normal metallic core of size $\xi$ with circulating currents around the vortex on the scale of the penetration depth $\lambda$. 
The density of vortices increases with $H$ and they begin to overlap with each other. Eventually, beyond a critical field strength $H_{c}$~~\cite{footnote1} the superconductor transitions into a metal in which the order parameter is suppressed to zero everywhere.

How does the superconductor evolve in the combined presence of disorder $V$ and magnetic field $H$? We address this question here specifically for a 2D sSC. We are motivated by several experimental puzzles: Firstly, it has been observed in many recent experiments~\cite{PhysRevLett.92.107005, STEINER200516, PRL98.127003, PRL101.157006, PhysRevB.77.140501, Ovadia2013} that the sheet resistance across a field-driven SIT at low $T$, shoots up by more than 8 orders of magnitude beyond a superconductor-insulator transition (SIT). Furthermore, the lineshape of magneto-resistance (MR) $\rho(H)$ is asymmetric~\cite{PhysRevLett.92.107005, PRL98.127003} with a sharp rise to a peak at $H_P$ followed by a gradual decrease to the normal state resistance~\cite{PRL98.127003, Shahar18}.
Interestingly, the MR-peak becomes sharper and stronger with increasing disorder in the films~\cite{STEINER200516, PRL98.127003,PRL99.257003, PRB91.220508R}. 

Several theoretical attempts have been made to explain this behavior ranging from Coulomb blockade on the SC-islands ~\cite{PhysRevB.73.054509}, to boson-localization~\cite{PhysRevLett.111.026801, EPL102.67008, PhysRevB.77.212501}, `superinsulator' and charge-vortex duality~\cite{Vinokur2008, Ovadia2013}. Suggestions have also been made that the high $B$ phase is a finite-temperature insulator -- akin to a many-body localized state~\cite{AnnalsofPhys321.1126}, however, no convergence has yet been reached. 

An second equally intriguing puzzle, which has received some attention, is the fate of the zero-bias peak (ZBP) in the local density of states (LDOS) of the vortex core region of the disordered SC-films in an orbital magnetic field. Such a peak in LDOS indicates an electronic bound state in the metallic cores of a clean vortex lattice, reminiscent of an Andreev bound state. This was first studied by Caroli, de Gennes and Matricon (CdGM)~\cite{CDGM_PhysLett9_307} for a single vortex in an otherwise uniform superconductor. The CdGM peak has been observed for periodic vortex lattices at low $H$~\cite{PRL64_2711, PRL101_166407}, however, in superconducting films that show  highly irregular array of vortices, presumably due to disorder, the CdGM peak is conspicuously absent~\cite{PRL96_097006, NatPhys11.332, PRB96_054509}. Instead, a series 
of resonances located at sub-gap energies have been found in the core region~\cite{PRL96_097006}, leaving a soft gap in density of states over the vortex region~\cite{PRB96_054509}.

In this paper, we obtain the phase diagram in the entire $H$-$V$ plane (see Fig.~\ref{fig:fig1}(b)). We provide a natural explanation for the strong MR-peak and its asymmetric shape, and for the absence of the CdGM peak in LDOS, solving two of the major long-standing puzzles for disordered superconducting films in a magnetic field.


\bigskip

\noindent {\it Model and methods} --
We describe a disordered type-II superconductor in the presence of orbital magnetic field by the attractive Hubbard Hamiltonian:
\beq
{\cal H}=-\sum_{\langle ij\rangle,\sigma} (te^{i\phi_{ij}}\hat{c}_{i \sigma}^{\dagger}\hat{c}_{j \sigma}+{\rm h.c.}) - U\sum_i \hat{n}_{i \uparrow}\hat{n}_{i \downarrow}\sum_{i\sigma} (V_i-\mu)\hat{n}_{i\sigma}
\eeq
Here, $t$ and $U$ denote the hopping amplitude and on-site attraction strength respectively, $c^\dagger_{i\sigma}$ ($c_{i\sigma}$) creates (annihilates) an electron on site $i$ with spin $\sigma$ on a two dimensional (2D) square lattice, and $n_{i\sigma}$ is the spin-resolved number operator for electrons on site $i$.  The orbital magnetic field is incorporated through the Peierls factor: $\phi_{ij}=\frac{\pi}{\phi_{0}}\int^j_i\mathbf{A}.d\mathbf{l}$, where $\phi_0=hc/2e$ is the superconducting flux quantum. We consider a uniform orbital field $\mathbf{H}=H \hat{z}$ and choose to work with the Landau gauge, $\mathbf{A}= Hx \hat{y}$. We use the model of a box disorder, $V_i\in [-V,V]$ chosen uniformly, to represent a homogeneously disordered system, and thus $V$ defines the strength of disorder.
The strength of attraction is denoted by $U$ and the average density $\rho=\sum_{i,\sigma}n_{i\sigma}$ is fixed by the chemical potential $\mu$. We have verified our key findings over a wide range of parameter space, and present results here for $U =1.2t$, and tune the chemical potential $\mu$ to fix the average density $\rho~(=\sum_{i,\sigma}n_{i\sigma})$ at $0.875$. It is believed that these parameters corresponds to a weak coupling sSC, with the coherence length $\xi \sim 12$ lattice spacing for $V=0, H=0$.  

In the clean ($V=0$) system, we take the advantage of the perfect periodicity of the vortex lattice by solving the eigenvalue problem for above ${\cal H}$ using a fully self-consistent Bogoliubov de-Gennes (BdG) method on a unit cell and then extending the wavefunction on a system made of typically $10\times 10$ unit cells using a repeated zone scheme (RZS)~\cite{PhysRevB.66.214502}.
We ensure that the total number of flux quantum is even in our full size system with periodic boundary condition.
All our calculations are performed at temperature $T=0$, and all energies are expressed in units of $t$.

We calculate two response functions -- the frequency-dependent conductivity $\sigma(\omega)$ and superfluid stiffness $D_s$ for our disordered superconductor using the Kubo formalism~\cite{PhysRevB.47.7995}, given by
\beq
\frac{D_s}{\pi} = \langle -k_x \rangle - \Lambda_{xx}(q_x=0,q_y\rightarrow 0,\omega=0)
\eeq
and
\beq
\sigma(\omega)=\frac {\langle -k_x \rangle - \Lambda_{xx}(\omega+i0^+)}{i(\omega+i0^+)}.
\eeq
From the above equations, we obtain the dissipative response, given by
\beq
\rm{Re}\sigma(\omega)= D_{s}\delta(\omega)+ Im\Lambda_{xx}(q=0,\omega)/\omega
\label{Sig_Ds}
\eeq
Here, $\langle k_x \rangle$ is the average kinetic energy along $x$-direction and $\Lambda_{xx}$ is the current-current correlation function~\cite{PhysRevB.47.7995}  along $x$-direction.

\bigskip

\begin{figure*}[ht]
{\includegraphics[width=15.5cm,keepaspectratio]{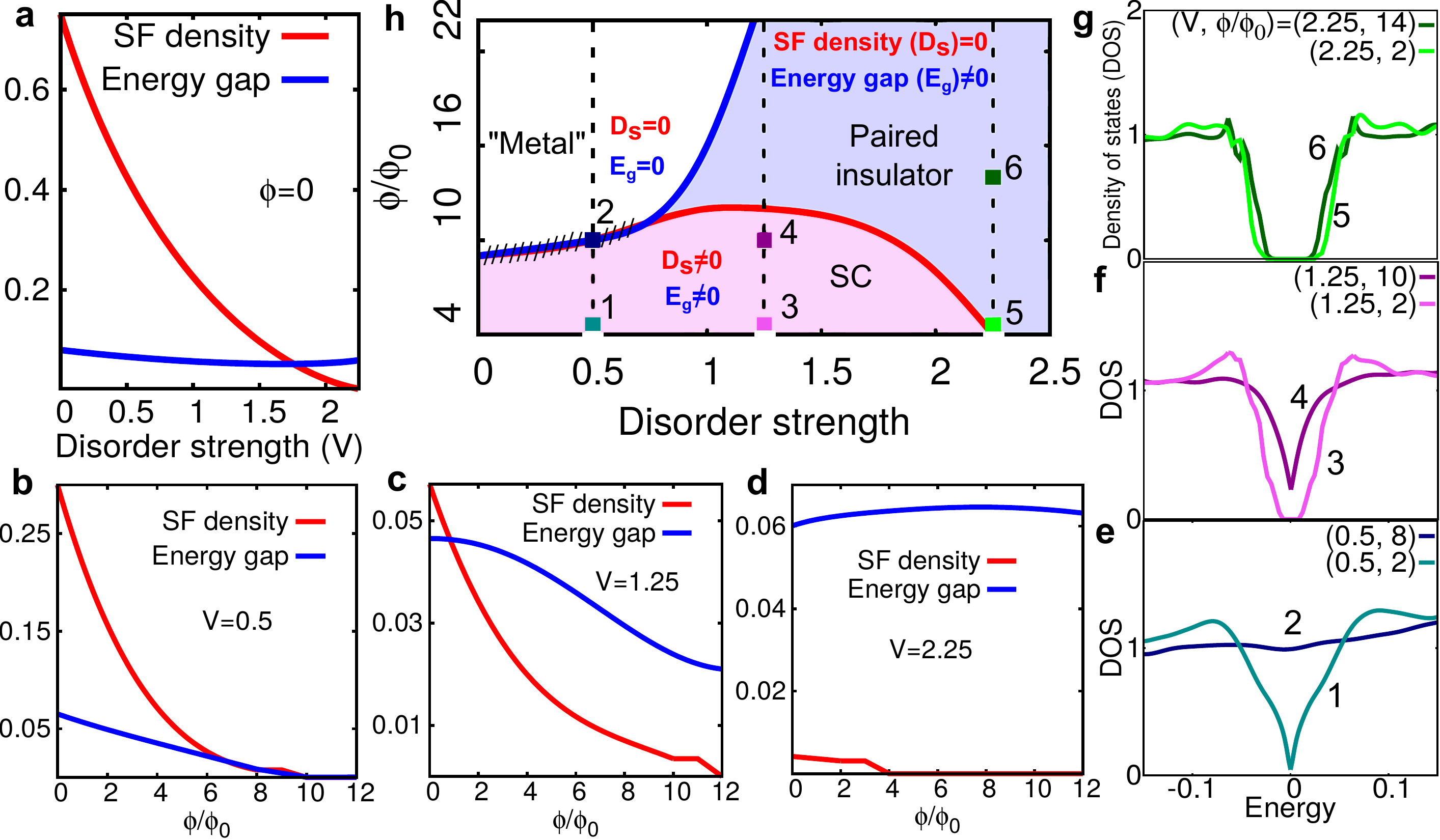}}
\caption{
(Color online) {\bf Phase diagram of the 2D superconductor in the disorder $V$ -magnetic field $H$ plane.} Panel (a) shows the evolution of the superfluid stiffness or superfluid density $D_s$ (in units of $t$) and the single particle spectral gap $E_g/t$ as a function of disorder $V/t$ for $H=0$. The vanishing of $D_s$ indicates a superconductor to non-superconductor transition at 
$V \approx 2.35$. $E_g$ survives beyond this transition. 
The dependence of $D_s$ and $E_g$ on $H$ are shown for disorder strengths: weak ($V=0.5$) [panel (b)], moderate ($V=1.25$) [panel (c)] and strong ($V=2.25$) [panel (d)]. The magnetic field $H$ is given in terms of $\phi$, the number of superconducting flux quanta penetrating a $36\times 36$ square magnetic unit cell in units of $\phi_0=h/(2e)$.
For weak disorder $D_s$ and $E_g$ collapse together at the critical $H_c$, however, for moderate and strong disorders, while $D_s$ decreases rapidly and vanishes at a critical $H_c$, $E_g$ is relatively unaffected by disorder and remains finite beyond $H_c$.  
Panels (e-g) portray the average DOS $N(\omega)$ at different values of $H$, at $V=0.5$, $1.25$ and $2.25$ respectively. The gap in the DOS rapidly fills up with increasing $H$ for weak disorder, but remains a hard gap for strong $V$.
We construct the phase diagram in the $V-H$ plane in panel (h). The phase boundaries: $H^{\rm D_s}_c$ (red trace) is  obtained from the vanishing of the superfluid stiffness $D_s$ and $H_c^{\rm E_g}$ (blue trace) from the vanishing of the spectral gap $E_g$. 
For $V=0$, the two critical fields coincide, consistent with the Abrikosov framework. For finite $V$, these two critical fields begin to separate out creating three distinct phases: 
(I) The region below the red trace is a true superconductor, in which both $D_s$ and $E_g$ are finite. (II) The region above the blue trace where both $D_s$ and $E_g$ are zero and the system is a metal or a gapless Anderson (fermionic) insulator. (III) The region between the red and blue traces, where $D_s$ vanishes, but $E_g$ remains finite, generating a novel pseudogapped insulator with Cooper-pairs.
}
\label{fig:fig2}
\end{figure*}

\noindent {\it Superfluid density and Energy gap:} --
The contrasting behavior of the superfluid stiffness $D_s$ and the single-particle energy gap $E_g$ with disorder $V$ at $H=0$, has previously been discussed with reference to Fig.~\ref{fig:fig1}; here we focus on their $H$-dependence shown in Fig.~\ref{fig:fig2} as a function of disorder. 
By comparing $D_s$ and $E_g$, we identify 3 disorder regimes: weak disorder around $V=0.5$ where $D_s \gg E_g$; moderate disorder around $V=1.25$ where $D_s \sim E_g$; and strong disorder around $V=2.25$ where $D_s \ll E_g$.

For weak disorder, (Fig.~\ref{fig:fig2}(b)) both $D_s$ and $E_g$ decrease with increasing $H$ and vanish at the critical $H_c$ (consistent with the expectations from Abrikosov theory for a clean sSC). The decrease of $D_s(H)$ close to $H_c$ shows a linear trend for weak disorder, which is consistent with the mean-field prediction within the Ginzburg-Landau formalism, as well as with experimental observations~\cite{NatPhys15_48}. 

For moderate disorder (Fig.~\ref{fig:fig2}(c)), on the other hand, $E_g$ decreases more gradually with $H$ compared to $D_s$; in fact the the two curves cross at a field corresponding to $\phi\approx \phi_0$, where $\phi$ is the magnetic flux through a unit cell and $\phi_0$ is the superconducting flux quantum. 

For high disorder (Fig.~\ref{fig:fig2}(d)), $E_g$ barely changes with increasing $H$, while $D_s$ declines precipitously (Fig.~\ref{fig:fig2}(b-d)). The energy gaps are extracted from the average density of states (DOS) $N(\omega)$ (Fig.~\ref{fig:fig2}(e-g)) shown for different combinations of $V$ and $H$. We notice that the standard BCS-type `hard' gap in $N(\omega)$ for an sSC turns into a soft pseudogap in a finite field $H$, particularly at weaker $V$. Extraction of $E_g$ from $N(\omega)$ featuring a soft gap is described in the supplementary materials. The DOS carries crucial information about superconducting correlations, and is discussed below. 

\bigskip
\noindent {\it Phase Diagram:} --
One of our central results is the discovery of an intermediate insulating region (between the superconductor and metal) as a function of $H$, for films with moderate to high disorder strengths. This insulating region is characterized by a finite $E_g$ but $D_s=0$, where the magnetic field generates, not Abrikosov vortices penetrating through superconducting regions, but core-less Josephson vortices formed from twisting phases in between superconducting `islands'. In particular, we show that the critical magnetic field $H_c$ required to suppress $D_s$, dubbed $H_c^{D_s}$, behaves rather differently at moderate to large $V$ from the critical field required to close the energy gap, dubbed $H_c^{E_g}$~\cite{fn}.

These findings are integrated in the $V$-$H$ phase diagram, Fig.~\ref{fig:fig2}(h). While the behavior of sSC along the two axes are well known, as explained above, our calculations presented here locate the phases in the entire parameter space. We depict the behavior of the two critical fields $H_c^{D_s}$ and $H_c^{E_g}$ as a function of disorder and find for $V \geq V_c$, the two $H_c$'s, branch out from each other; for smaller $V$ any differences between these two critical fields is not discernible from our calculations. The $V$-$H$ plane shows three distinct phases: (i) Superconductor, where both $D_s$ and $E_g$ are finite. (ii) Metal, where both $H_{c}^{D_s}$ and $H_c^{E_g}$ disappear~\cite{fn3}. (iii) Gapped and paired insulator, where $D_s$ is zero but $E_g$ remains finite.

Based on our phase diagram, we see that a direct transition from a superconductor to a metal, with increasing magnetic field, as occurs in the Abrikosov description~\cite{AVL} of clean type-II superconductors, is not found for larger disorder strengths. 
Instead an insulating state of pairs intervenes in which the resistance diverges as the temperature approaches zero. We thus provide a new paradigm for the fate of superconductivity as a function of magnetic field for films with stronger disorder.
\begin{figure}[h]
{\includegraphics[width=8.5cm,keepaspectratio]{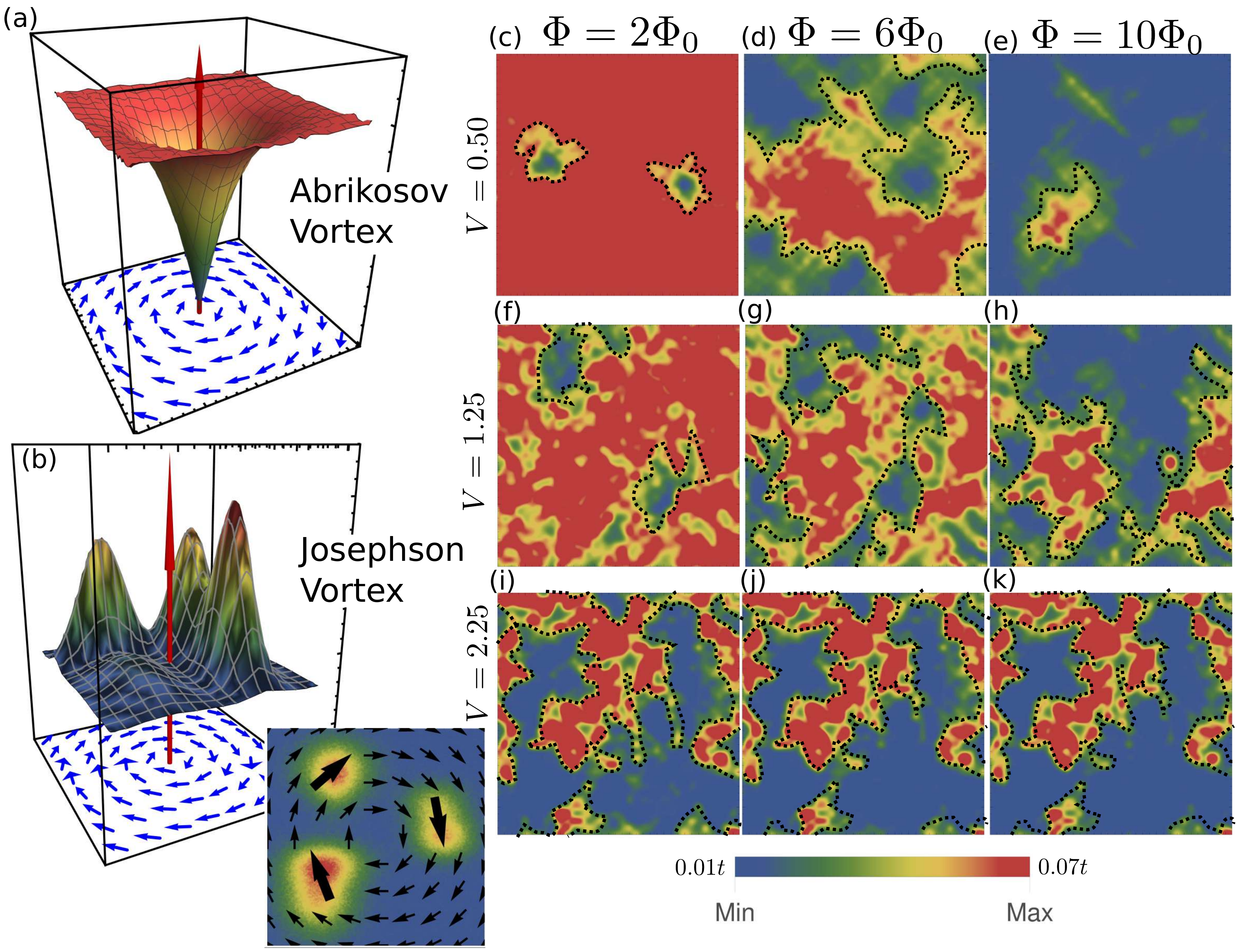}}
\caption{(Color online) {\bf Abrikosov vortices transform to Josephson vortices with increasing disorder} \\
Panels (a): Schematic plot of an Abrikosov vortex showing the suppression of the pairing amplitude at the vortex core and the curling of the phase of SC order parameter around the core-center depicted as the projection on the floor.
Panel (b): Schematic plot of a Josephson vortex with its core in the non-SC region (zero pairing amplitude) surrounded by 3 SC-islands (finite pairing amplitude $\Delta$). The phases are largely uniform on the islands and twist in the non-superconducting regions in between.
The red arrows in (a) and (b) represent the magnetic field threading the system.
BdG calculations give the spatial map of the superconducting pairing amplitude $\Delta({\bf r})$ with $H$ increasing from left to right along each row for weak $V = 0.5$ (along top rows – panel (c-e)), moderate $V = 1.25$ (along middle rows – panels (f-h)) and strong $V = 2.25$ (along bottom rows – (i-k)) disorder strengths.
For weak $V=0.5$, the flux lines thread through regions of weaker $\Delta$, punching out increasing number of holes of size $\sim \xi$, with increasing $H$. Their eventual overlap collapses $\Delta$ everywhere and hence destroys superconductivity in the system for $H > H_c$.  
In contrast, flux lines are accommodated in regions where $\Delta\approx 0$ in the spatially  inhomogeneous landscape of $\Delta({\bf r})$ induced by strong disorder $V=2.25$ (panels (i-k)) They, therefore, do not produce any additional holes on the SC-islands and thus the $\Delta$-profile remains insensitive to $H$. The evolution for moderate disorder $V=1.25$ is depicted in panels (f-h).
}
\label{fig:fig3}
\end{figure}

\bigskip

\noindent {\it Local pairing amplitude: location of vortices:} --
We next discuss two important questions: What is the underlying mechanism of the field driven transition of a type-II disordered superconductor to a non-superconducting state? And what is the nature of the resulting non-superconducting state?
By analyzing the field induced spatial inhomogeneities in the pairing amplitude and the nature of vortex cores for different disorder strengths, we obtain useful insights that help answer these questions.

Fig.~\ref{fig:fig3} shows the evolution of vortices from an Abrikosov vortex with a metallic core for low disorder to a ``core-less" Josephson vortex at higher disorder~\cite{PhilMag86.3569}.
For low disorder, the pairing amplitude is rather homogeneous and when vortices form
they penetrate the superconducting region and dig a hole where the pairing amplitude $\Delta({\bf r})$ vanishes creating a metallic core (Abrikosov vortex) Fig.~\ref{fig:fig3}(c-e). 
With increasing disorder, Fig.~\ref{fig:fig3}(i-k) the pairing amplitude $\Delta({\bf r})$ becomes strongly inhomogeneous even in the absence of $H$ and form SC puddles separated from insulating regions. It is now energetically favorable for the flux-lines to penetrate the system in regions where $\Delta$ is already low typically coinciding with high disorder regions. These Josephson vortices cause the SC phase on the adjacent grains to wind around.  
As $H$ increases, the density of flux-lines increase, causing greater winding of the phases on the surrounding puddles, ultimately destroying SC at $H=H_c$. Henceforth, we will use $H=H_c$ in place of $H_c^{D_s}$ for notational simplicity.

\begin{figure}[h]
{\includegraphics[width=8.5cm,keepaspectratio]{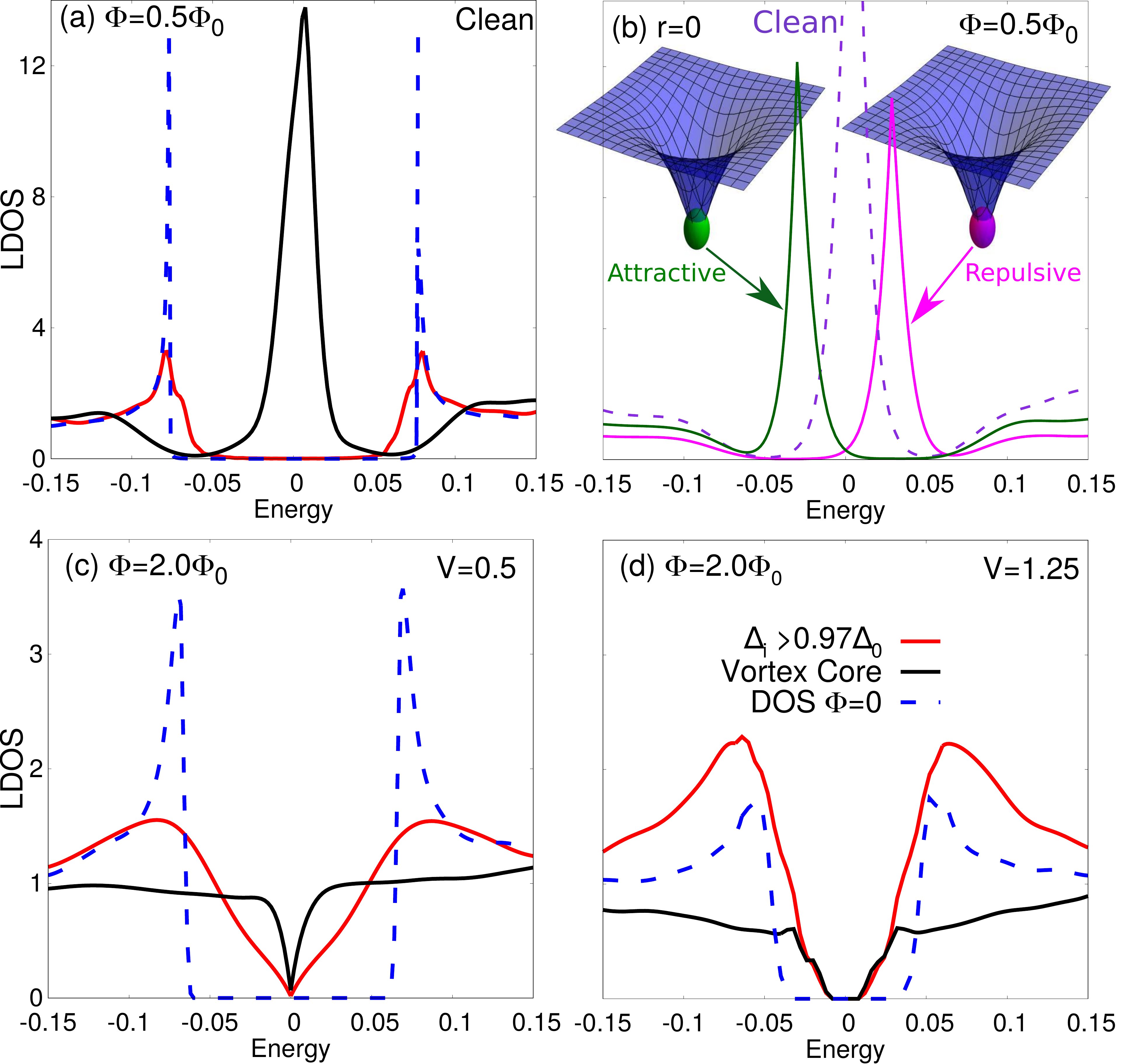}}
\caption{
(Color online) {\bf Fate of zero bias peak in the presence of magnetic field and disorder:}
Panel (a) shows the LDOS in the core region $N({\bf r}_v, \omega)$ (averaged over the vortex center, nearest and next nearest neighboring sites, of a clean system (black), which depicts the CdGM peak near $\omega\approx 0$. The LDOS averaged over sites far away from the vortex center (red) where $\Delta({\bf r})$ is large, and the clean BCS-DOS (broken blue trace) are also shown for comparison.
Panel (b) depicts the shift of the CdGM-peak
(dotted) to $\omega = \pm \omega_0\propto V_0$ when a single impurity of strength $\pm V_0$ is placed at the center of the vortex. 
Panel (c) and (d) show the LDOS similar to panel (a), but for $V=0.5$ and $V=1.25$ respectively, where we discover that the CdGM peak is dramatically absent! It gets replaced by a gap-like feature around $\omega \approx 0$. The width of this gap is larger for higher $V$. The weakening coherence peaks with increasing disorder for $H=0$ is evident in these panels. 
}
\label{fig:fig4}
\end{figure}

\begin{figure*}[ht]
{\includegraphics[width=16.0cm,keepaspectratio]{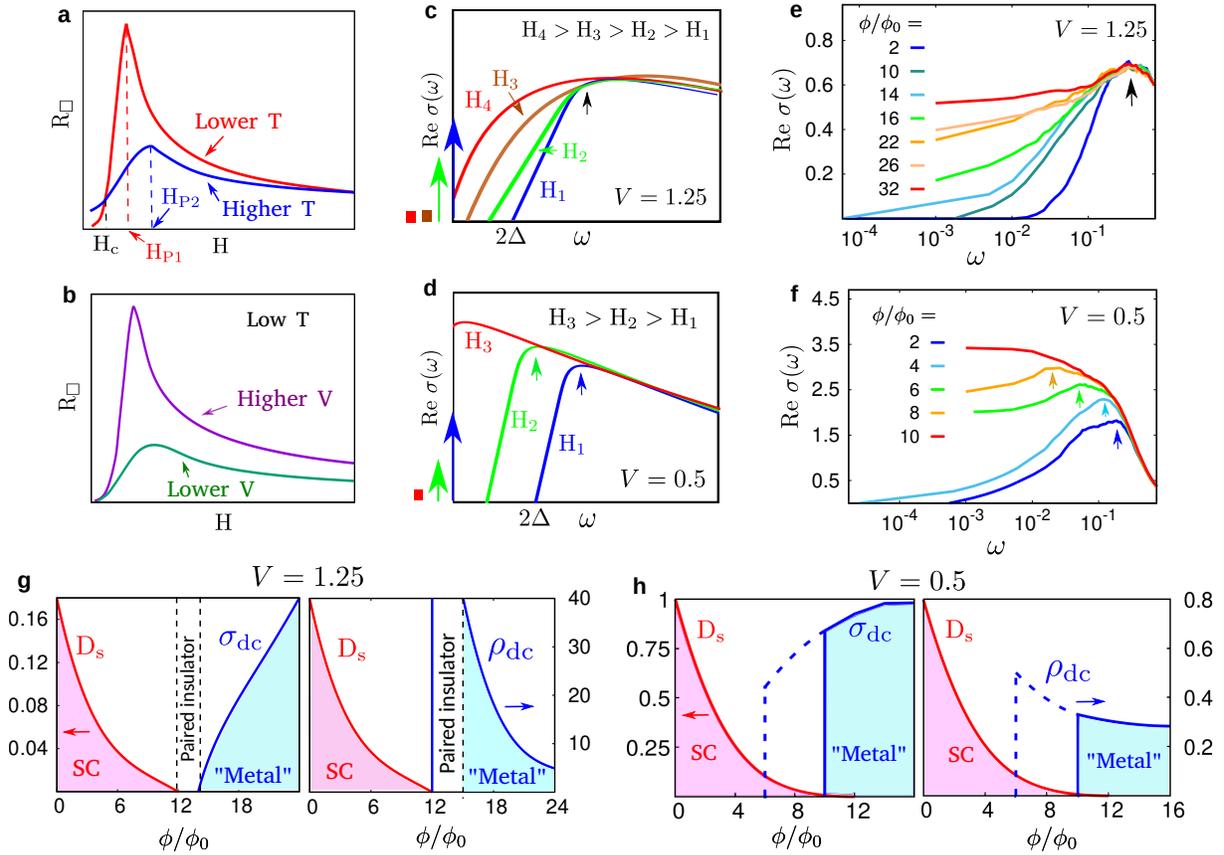}}
\caption{
(Color online) {\bf Origin of colossal magneto-resistance (MR) peak in disordered superconducting films.}
Panel (a) shows a schematic of the experimental sheet resistance $R_{\square}$ of a typical disordered SC-film as a function of an applied magnetic field $H$. $R_{\square}$ rises rapidly by many orders of magnitude and reaches a peak at $H=H_P$ then decays to the its normal state value gradually for $H \gtrsim H_P$. The MR-peak becomes sharper and its position $H_P$ marches down to lower $H$ with decreasing of $T$. Panel (b) depicts the experimental magnetoresistance peak that becomes intense and sharper with increasing $V$ at low temperature. 
Panel (c) shows frequency-dependent conductivity $\sigma(\omega)$ obtained from BdG calculations and Kubo formalism for a moderately disordered ($V=1.25$) 2D-sSC for different $H$. For low $H$, $\sigma(\omega)=0$ below a gap $\omega^{\ast}$ and increases linearly $\sigma(\omega)\approx (\omega-\omega^\ast)$ above the gap.
The gap $\omega^\ast$ decreases with $H$, disappearing at $\phi^{\ast} \approx 14\phi_0$. Interestingly, the field $H^{\ast}$ at which the superfluid response vanishes is $\approx H_c$, implying that the gap in $\sigma(\omega)$ tracks $D_s$, and not $E_g$. 
For $H \rightarrow H^{\ast}$, the gap closes and $\sigma(\omega) \sim \omega^{0.4}$ becomes sub-linear, and for larger $H \geq H^{\ast}$, $\sigma(\omega \rightarrow 0)$ is finite.
For all $H$, $\sigma(\omega\rightarrow \infty)\sim \omega^{-2}$ yielding the standard Drude tail (not shown here).
Panel (d) shows $\sigma(\omega)$ for weak disorder ($V=0.5$). The gap gets filled by moving the peak and shifting weight of $\sigma(\omega)$ to lower $\omega$ progressively with $H$.
Panel (e) presents BdG results for $V=1.25$ corresponding to panel (b) on a semi-log scale, which emphasizes that the gap-filling of $\sigma(\omega)$ with increasing $H$ occurs by shifting weight from higher to lower frequency; the peak position of $\sigma(\omega)$ does not alter with $H$.
Panel (f) presents BdG calculations of  $\sigma(\omega)$ on a semi-log scale, corresponding to panel (e) for weak disorder $V=0.5$ showing a downshift of the peak position of $\sigma(\omega)$ indicated by arrows.
Panels (e) and (f) highlight the contrast in the mechanism of gap filling for weak and moderate disorder strengths.
Panel (g) shows $D_s$ (red) and $\sigma_{dc}$ (blue) as well as sheet resistance $\rho_{dc}$ (blue) in the subfigures for moderate disorder as functions of $H$, which demarcate the SC (pink region) and metal (blue region). Between these two regions, there exists a gapped-insulating phase (in the limit of $T\rightarrow 0$) in a window of $H$. This window of insulating behavior is a direct consequence of the ``pseudogapped" region in the phase diagram of Fig.~2(h). Note that, upon depletion of superfluidity for $\phi_c\approx 12\phi_0$, the resistivity $\rho_{dc}(12\phi_0)$ shows a sharp rise, tantalizingly similar to the experimental signature represented in panel (a).
Panel (h) shows the corresponding evolution of $\sigma_{\rm dc}$ and $\rho_{dc}$ along with $D_s$ for weak disorder $V=0.5$, where the pseudogap is absent.
Not only does the thin insulating regime between SC and metallic phase of panel (g) disappears, but the SC short circuits the overlapping SC and metallic regions obtained from self-consistent calculations.
A finite $\rho_{dc}$ appears shown by the solid blue trace on the right half of panel (h), with no divergence, only gradually increasing causing only a hump in the magnetoresistance. This contrast of responses between films of higher and lower disorder content, also finds experimental support, as schematically represented in panel (b).
}
\label{fig:fig5}
\end{figure*}

\bigskip

\noindent {\it Local density of states (LDOS) $N({\bf r},\omega)$ at the vortex core:} --
Because the cores of a clean vortex lattice are metallic, they support states at $\omega \approx 0$. In fact, Caroli, de Gennes and Matricon (CdGM)~\cite{CDGM_PhysLett9_307} showed that a simple model of a single vortex with normal metallic core is sufficient to produce experimentally observed sharp conductance peaks near zero bias in the LDOS $N({\bf r}_v, \omega)$, indicative of low lying bound states around the core ${\bf r}_v$~\cite{fn4}. The behavior of the LDOS at the core of a clean vortex lattice with a sparse array of vortices for weak field $H$ is shown in Fig.~\ref{fig:fig4}(a). 

In contrast, a disordered system Fig.~\ref{fig:fig4}(b,c) shows the absence of the CdGM peak in $N({\bf r}_v, \omega)$, featuring a gap whose width increases with disorder. The appearance of this gap in the presence of disorder turns the vortex cores insulating, quite unlike the metallic cores of Abrikosov lattices. The reason behind this change of character relies on the underlying non-superconducting state from which Josephson-type vortices are created. Such Josephson vortices typically nucleate in regions of large disorder fluctuations where the pairing amplitude is suppressed.
The low energy states of the disordered system is already consumed by the SC-islands. This explains the gapped (and hence insulating) nature of the cores, at least for large $V$.

 We have also found that the presence of a single impurity at the vortex-center diminishes the intensity of the resonances in the LDOS and shifts them to a finite $\pm \omega_0$, where the sign of the shift depends on the sign of the impurity potential and its magnitude is proportional to the strength of the impurity (Fig.~\ref{fig:fig4}(b)). In fact, the shift of resonances at and around the vortex center due to the presence of impurities explains the emergent gap in $N({\bf r}_v)$ (as seen in panels (c) and (d)) when the samples are homogeneously disordered. More details are provided in the supplementary materials.

\medskip

\noindent {\it Origin of gigantic magnetoresistance} --
We next turn towards providing an explanation for the sharp MR-peak observed beyond SIT in disordered superconductors, as recorded by many experiments~\cite{PhysRevLett.92.107005, STEINER200516, PhysRevB.77.140501}. 

For the clean sSC as expected we find no absorption at $H=0$ for $\omega \leq 2\Delta_0$, where $\Delta_0$ is the single particle gap. Notice from Eq.~(\ref{Sig_Ds}) that the superfluid stiffness $D_s$ is the weight of the $\delta$-function at $\omega=0$ in $\rm{Re}[\sigma(\omega)]$ but being a lossless conduction term, it does not contribute to MR.
Moderate disorder ($V=1.25$) generates states within the gap and shifts spectral weight toward lower frequencies (see Fig.~\ref{fig:fig5}(b)). We find that $\sigma(\omega)$ depends linearly on $\omega$ beyond the gap for small $H$. With increasing $H$, the size of the gap shrinks as the low lying states get progressively filled up, and results in a non-linear $\omega$-dependence of $\sigma(\omega)$. We find that the gap in $\sigma(\omega, H \neq 0)$ is consistently smaller than $E_g$ extracted from the average DOS $N(\omega)$, for all disorder strengths studied here the difference between the two gaps rises with disorder strength. 

Interestingly, the position of the broad peak in $\sigma(\omega)$ at $\omega=2\Delta_0$ for $V\rightarrow 0$ does not change with $H$ for moderate disorder ($V=1.25$), even though the (low-$\omega$) tail of $\sigma(\omega)$ gathers more weight with increasing $H$. In contrast, the sharper peak in $\sigma(\omega)$ for weak disorder ($V=0.5$) marches down to lower $\omega$ in addition to developing a weaker tail toward low-lying states. So a combined effect of the downshift of the peak, as well as strengthening of low-$\omega$ tail become responsible for the filling of the gap in $\sigma(\omega)$ for weak disorder, whereas such downshift of the peak does not occur for enhanced $V$ (Figs.~\ref{fig:fig5}(c,d)), resulting in a qualitatively different nature of gap- filling in $\sigma(\omega)$.

We map out the superconducting and metallic phases based on the behavior of the superfluid phase stiffness $D_s$ and the conductivity $\sigma_{\rm dc}$ as a function of magnetic field for various disorder strengths (see the insets of Fig.~\ref{fig:fig5}(e,f)). $\sigma_{\rm dc}$ is defined as the limit $\sigma(\omega \rightarrow 0)$ for a given $V$ and $H$, extracted from Fig.~\ref{fig:fig5}(b).
We observe two qualitatively different behavior: For moderate disorder ($V=1.25$ in panel (e)), $D_s$ decreases with increasing field and vanishes at $H_{\rm c}$. However, dissipative response only appears for $H>{\tilde H}_{\rm c}$ where ${\tilde H}_{\rm c} > H_{\rm c}$. In the region between these two critical fields the system remains as a paired insulator in which both $D_s$ and $\sigma_{\rm dc}$ are zero. On the other hand, for weaker disorder ($V=0.5$ in panel(f)), $D_s$ remains finite at $\phi \approx 6\phi_0$, whereas $\sigma_{\rm dc}$ jumps up to a finite value from zero. However, the dissipative response is short-circuited by the finite $D_s$, ensuring dissipationless (superfluid) transport in the system. Resistive transport with finite $\sigma_{\rm dc}$ sets in for $H \geq H_c$ only after the collapse of $D_s$. We emphasize that the jump of $\sigma_{\rm dc}$ at $\phi_c/\phi_0 \approx 6$ occurs because the peak of $\sigma(\omega)$ (Shown in Fig.~\ref{fig:fig5}(d)) marches down in $\omega$ with increasing $H$, which distinguishes it qualitatively from its behavior at stronger disorder.

In order to compare with experiments, we show the behavior of the resistivity $\rho_{\Box}$ (Figs.~\ref{fig:fig5}(e,f)) obtained from the inverse of $\sigma_{\rm dc}$
For moderate disorder ($V=1.25$), because $D_s$ collapses before the onset of $\sigma_{\rm dc}$, as $H$ increases, there is a window of 
magnetic field between ${\tilde H}_{\rm c} < H < H_{\rm c}$,
where the resistivity truly diverges. On the other hand, $\rho_{\Box}=0$ for $H < {\tilde H}_{\rm c}$, develops a sharp peak at ${\tilde H}_{\rm c}$ and decreases gradually beyond $H_{\rm c}$.
We note that in a recent experiment~\cite{Shahar18} it is claimed that the MR peak position 
$H_{\rm P} \rightarrow H_{\rm c}$
as $T \rightarrow 0$, an observation consistent with our $\rho_{\Box}$ in panel (e).
On the other hand, for weak disorder ($V=0.5$), $\rho_{\Box}$ remains finite in the metal or is zero in the SC phase and leads to only a very weak hump in MR. Consistent with this picture, we see in Fig.~\ref{fig:fig5}(f) that for such low disorder region, the spectral gap collapses at the same critical field where $D_s$ vanishes.

Thus, our analysis above, not only explains the peak in the MR, it also explains its highly asymmetric shape (a sharp rise followed by a gradual fall), in agreement with experiments~\cite{PhysRevLett.92.107005}. 

{\it Conclusion.} -- 
We have provided a natural explanation of the origin of the MR-peak as arising from the emergent granularity of the local order parameter in combined presence of $V$ and $H$. One of our main results is the prediction of a paired insulating phase beyond a critical disorder in a superconductor. This phase has a hard gap at zero field, develops a pseudogap in a finite field, and persists up to large fields, well beyond the critical field at which the superfluid density vanishes. While our results clearly show that there is a single particle gap or pseudogap, an open question remains about the nature of transport by pairs or collective modes in the pseudogap phase: is it metallic with a finite dc conductivity or insulating?
Our results show that, within the uncertainties of our simulations, the gap measured in the absorption threshold of $\sigma(\omega)$ is less`\cite{Sherman2015,PhysRevB.93.180511} compared to twice the single particle energy gap with increasing field. These suggest that a possible Bose metal to Bose insulator transition in the pseudogap  region should be explored further.
\medskip

\noindent {\it Acknowledgment:} -- We thank Pratap Raychaudhuri for valuable discussions.

\bibliographystyle{apsrev4-1}
\bibliography{Draft}

\begin{thebibliography}{51}
\bibitem [{foo()}]{footnote1}%
  \BibitemOpen
  \href@noop {} {}\bibinfo {note} {In our study we consider a strongly type-II
  superconductor so first critical field for the insertion of a single vortex
  $H_{c1}\rightarrow 0$. For notational simplicity $H_c$ refers to $H_{c2}$ in
  standard literature.}\BibitemShut {Stop}%
\bibitem [{fn()}]{fn}%
  \BibitemOpen
  \href@noop {} {}\bibinfo {note} {While we introduced in text the two critical
  fields $H_c^{D_s}$ and $H_c^{E_g}$ for clarity, in true sense the `critical
  field' is identified by $H_c^{D_s}$ that marks the destruction of type-II
  superconductivity. For all our discussions, the simplified notation $H_c$
  always represents $H_c^{D_s}$.}\BibitemShut {Stop}%
\bibitem [{fn3()}]{fn3}%
  \BibitemOpen
  \href@noop {} {}\bibinfo {note} {{within our numerics, we cannot distinguish
  with confidence the ``metallic" phase from an Anderson insulator. Our
  identification of a ``metallic" phase relies on the fact that the
  localization length, particularly for small disorder, remains larger than our
  system size}}\BibitemShut {NoStop}%
\end{thebibliography}

\pagebreak

\onecolumngrid
\vspace{\columnsep}
\begin{center}
\textbf{\large Supplementary material for ``New paradigm for a disordered superconductor in a magnetic field''}
\end{center}
\vspace{\columnsep}
\twocolumngrid

\setcounter{equation}{0}
\setcounter{figure}{0}
\setcounter{table}{0}
\setcounter{page}{1}
\setcounter{enumiv}{0}
\makeatletter
\renewcommand{\theequation}{S\arabic{equation}}
\renewcommand{\thefigure}{S\arabic{figure}}
\renewcommand{\bibnumfmt}[1]{[S#1]}
\renewcommand{\citenumfont}[1]{S#1}

\section{Extraction of the energy gap $E_{\rm g}$}\label{Eg}
\begin{figure}[h]
\includegraphics[width=8.5cm,height=10.0cm,keepaspectratio]{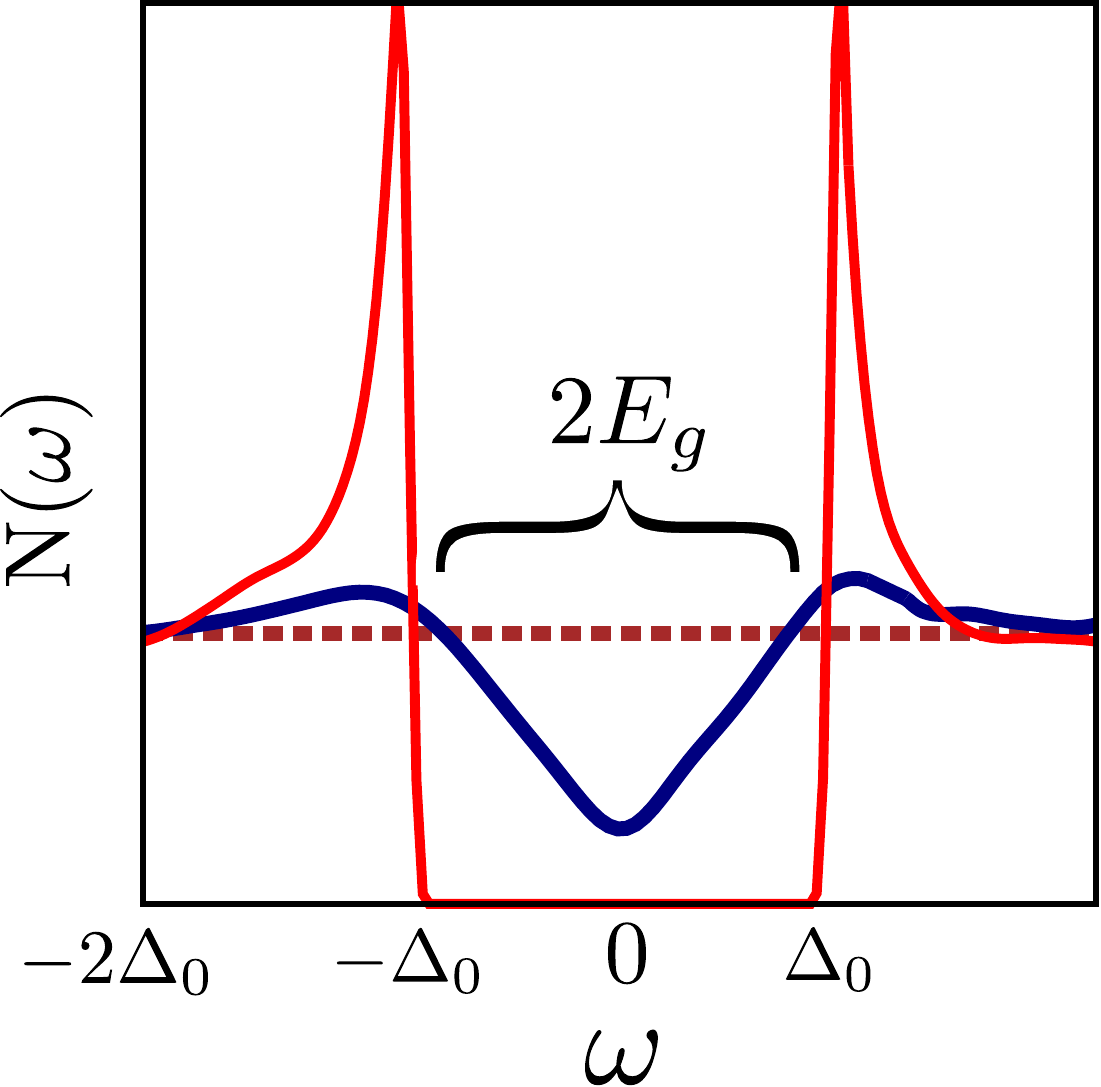}
\caption{Definition of $E_{\rm g}$ when the hard spectral gap of a clean sSC (red trace) turns into a soft gap (blue trace) due to the combined presence of disorder $V$ and magnetic field $H$. $E_{\rm g}$ is extracted by defining the dashed baseline as described in the text.
}
\label{fig:SI1}
\end{figure}

The single particle energy gap $E_{\rm g}$, as presented in Fig.~\ref{fig:fig2} of the main text, is extracted from the average density of states (DOS) $N (\omega)$. The energy spectrum of a clean s-wave superconductor (sSC) at $T=0$ in the absence of a  magnetic field features a hard gap $\Delta_0$, whose magnitude equals the homogeneous pairing amplitude, as shown by the red trace in Fig.~\ref{fig:SI1}. However, the simultaneous presence of $V$, and in particular, $H$, softens the gap in a manner depicted by the blue trace in Fig.~\ref{fig:SI1}. Naturally, many of the traces of $N (\omega)$ included in Fig.~\ref{fig:fig2}(e, f, g) show similar traces. While it is easy to define a hard gap objectively, identification of the gap-scale for a soft gap needs to be defined, and we proceed here to describe our method of identification of the gap-scale.
We construct a horizontal dashed line parallel to $\omega$-axis along $N(\omega)$-axis, such that it matches the profile of $N(\omega)$ at $\omega=-2\Delta_0$, where $\Delta_0$ is the clean BCS energy gap.
The trace of $N(\omega)$ changes rather gradually beyond this cut-off energy of $2\Delta_0$, we verified that for all smaller $\omega$ up to the lower band edge, $N (\omega)$ becomes more or less $\omega$-independent.
We choose not to use such matching of the baseline and $N(\omega)$ on the positive bias because the presence of the van Hove singularity of the underlying tight binding model obscures a good match. Finally, $E_{\rm g}$ is extracted from the intersection of the horizontal line with the low-energy part of $N(\omega)$ curve, as illustrated in Fig.~\ref{fig:SI1}.
Although such a definition of $E_{\rm g}$ would be somewhat ad hoc, the trend of resulting $E_{\rm g}$ as a function of a tuning parameter is expected to represent physical situation adequately, and similar ideas are often employed in experiments~\cite{HermannSuderow}.

\section{DC resistivity $\rho_{\rm dc}$ in the units of $\hbar/e^{2}$}

In our theoretical calculation we use the ``box"-model of disorder parameterized in terms of the single microscopic parameter $V$. While this is theoretically satisfactory, experiments have no access to such a parametrization. In order to establish a connection with experimentally accessible values, we develop a correspondence between the disorder strength $V$ in units of the hopping amplitude and the DC resistivity $\rho_{\rm dc}$ expressed in units of $\hbar/e^{2}$.
 
In order to express $\rho_{\rm dc}$ in the units of $\hbar/e^{2}$, we first focus calculate the relaxation time $\tau$ of electrons in the presence of disorder~\cite{smith1989transport_SI}. 
We next express the Drude conductvity as:
\begin{equation}
\sigma_{\rm dc}=\frac{\langle n\rangle e^{2}\tau}{m^{\ast}}= \langle n\rangle \frac{v_{f}\tau}{k_{f}} \frac{e^{2}}{\hbar}~.
\label{Eq:DrudeSig}
\end{equation}
Here, $\langle n \rangle$ is the average electron density, $m^{\ast}$ is the effective electron mass, $k_{f}$ and $v_{f}$ are the magnitude of Fermi wave-vector and Fermi velocity respectively. The Fermi velocity $v_{f}$ is obtained from the momentum gradient of the energy band of the underlying tight-binding model. For our estimates, $v_{f}$ is averaged over the Fermi surface contour $\mathbf{k_{f}}$ corresponding to $\epsilon_{k_f}=\mu$.  

In the table below we present results for the DC resistivity $\rho_{\rm dc}=1/\sigma_{\rm dc}$ in units of $\hbar/e^{2}$, and the dimensionless parameter $1/k_{f}l$ for each disorder strength, where $l=v_{f}\tau$ is the mean free path of electrons.

\begin{center}
    \begin{tabular}{c|c|c} 
    \hline
      \textbf{V} (in the unit of $t$) & \textbf{$\rho_{\rm dc}$} (in the unit of $\hbar/e^{2}$) & \textbf{$1/k_{f}l$}\\
      \hline
      0.5 & 0.32 & 0.05\\
      \hline
      1 & 1.28 & 0.2\\
      \hline
      1.25 & 1.99 & 0.31\\
       \hline
      2 & 5.10 & 0.8\\
      \hline
      2.25 & 6.45 & 1.01\\
    \end{tabular}
\end{center}
\begin{figure}[h]
\includegraphics[width=8.5cm,height=10.0cm,keepaspectratio]{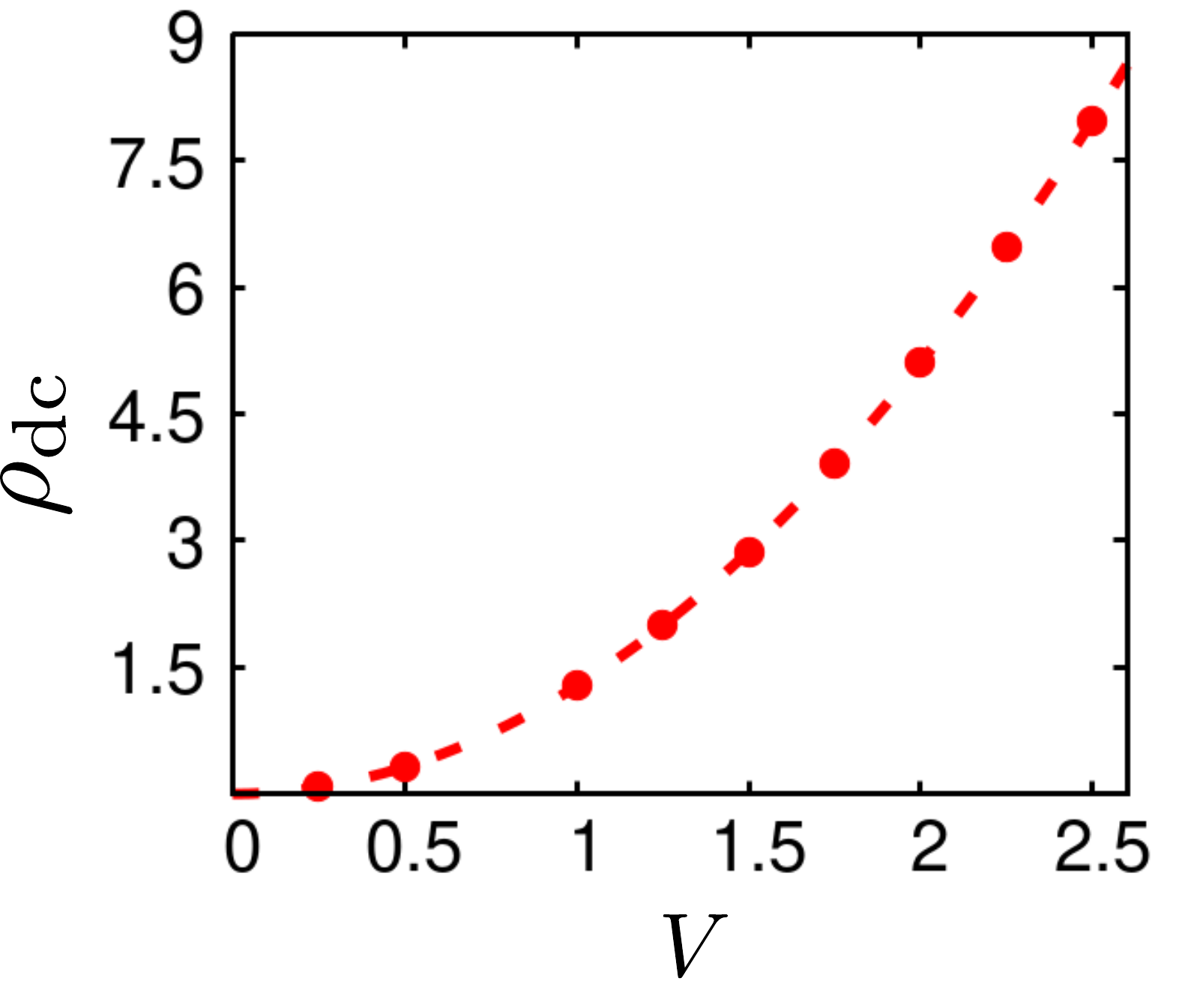}
\caption{Variation of the DC resistivity $\rho_{\rm dc}$ (in the unit of $\hbar/e^{2}$) with respect to disorder strengths $V$.  $\rho_{\rm dc}$ exhibits a quadratic pattern as a function of $V$, as depicted by the dashed line, behaving as $\approx 1.3 V^{2}$. }
\label{fig:rhoV}
\end{figure}

We also plot $\rho_{\rm dc}$ versus $V$ using the tabulated values in Fig.~\ref{fig:rhoV}, where the dashed (best fit) follows a quadratic behavior. This is expected because $\rho_{\rm dc} \sim \tau^{-1}$ and $\tau^{-1} \propto \langle V^2 \rangle$ according to Fermi golden rule estimates. The constant of proportionality depends on the details of model parameters. We also draw attention of the readers to our observation that the estimate of $V_c(H=0)$ corresponds to $k_{f}l \approx 1$. Finally, we note that our evaluation of $\rho_{\rm dc}$, as tabulated here matches favorably with the values for superconducting films~\cite{PRL98.127003_SI,Ovadia2013_SI,STEINER200516_SI}.

\section{Fate of CdGM Peak in the presence of impurities}
\label{ExplainDV}

We focus below on the self-consistent BdG calculations for a few impurities at or near vortex center. 

\begin{figure*}[ht]
 \centering
    {\includegraphics[width=15.0cm]{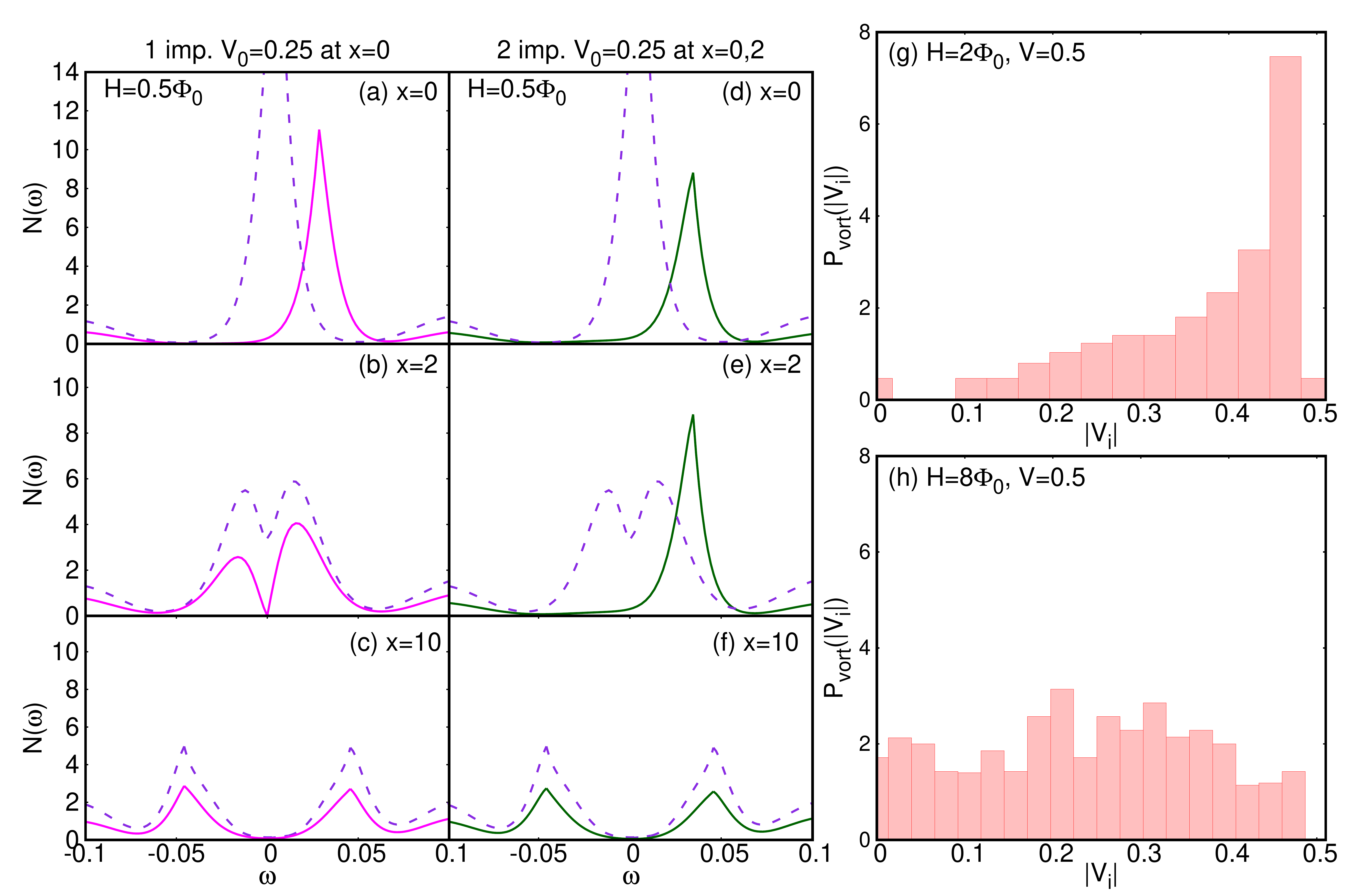}}
        \caption{{\bf{Local density of states (LDOS) around vortex cores:}} (a-c) LDOS around the vortex center ($x=0$) for single impurity at the vortex core of strength $V_0=0.25t$. (d-f) LDOS for two impurities, one at the vortex core and other at $x=2$ two sites away from the vortex core in the $x$-direction. The presence of the repulsive impurity shifts the CdGM peak to positive energies and makes it weaker. Distribution of disorder strength at the vortex core for disorder strengths $V=0.5$ for $H=2 \Phi_0$ (panel g), and for $H=8 \Phi_0$ (panel (h)).}
    \label{fig:SI2}
\end{figure*}

First, we obtain the spectrum for a single impurity of strength $V_0=0.25$ at the vortex center from our self-consistent BdG numerics. We find that the vortex gets pinned to the single impurity, 
The LDOS for such a single impurity at the vortex core $x=0$ is shown in Fig.~\ref{fig:SI2}(a) (thick trace) and compared to that of a clean system (dotted trace). We find the CdGM peak moves away from $\omega \approx 0$ to a positive energy $\omega \approx +\omega_0$. The intensity of the peak reduces as well compared to the clean case.
Away from the core, the effect of this impurity wears off, as seen from Fig.~\ref{fig:SI2}(b), and by a distance of 10 sites away from the center, the LDOS closely resembles that of the clean system as seen from Fig.~\ref{fig:SI2}(c).

We next solve the problem of two impurities of identical strength $V_0=0.25$, one set at the vortex core $x=0$, and the other located at $x=2$ along the x-direction. The presence of the second impurity forces the LDOS at center to move to positive energies as before with a reduced intensity, as shown in Fig.~\ref{fig:SI2}(d). Also, the split peak DOS feature at $x=2$ found in Fig.~\ref{fig:SI2}(b) moves to a single peak also at $\omega \approx +\omega_0$. Far away from the center, the LDOS recovers the features of the clean system. 

Similar effects are also checked for the attractive impurities with broadly similar findings except for the shift of CdGM peaks now towards the negative energy $\omega \approx -\omega_0$. This can be easily understood. A repulsive impurity reduces the local electron density, and as a result, it becomes difficult to extract electrons from those sites. Therefore, the LDOS shows a shift to positive(negative) energies as the repulsive(attractive) impurity forces the weight of hole(electron) or $u^2(v^2)$ larger at that site. Previous self-consistent investigation of point-impurity at the vortex core regions also reveals similar shifting of the CdGM peaks at higher energies~\cite{HanPRB_SI}.

The calculations mentioned above provide useful insights on the effects of impurities on core states. The observed shift of CdGM-type peaks, on an average, gives rise to a soft gap when the LDOS is averaged over a region surrounding the vortex center. Such series 
of resonances located at sub-gap energies is detected in the STM mesurements near the core region~\cite{PRL96_097006_SI}, leaving a soft gap in density of states over the vortex region~\cite{PRB96_054509_SI}.

The natural question arises then: What is the strength of the impurities at the vortex cores for a uniformly disordered superconductor in a magnetic field? 
In Fig.~\ref{fig:SI2}(e) we present the distribution of the local impurity potentials at the vortex core for low magnetic field strengths obtained from our BdG studies. The distribution is peaked toward higher disorder strengths. Thus for low-field, the flux-lines prefer to penetrate through the regions of significant disorder fluctuations. This is natural, as the vortices thread through areas where pairing amplitude is already weak to minimize pairing energy cost. 

With increasing $H$, the number of vortices increases, and they cannot all be accommodated in the strongly disordered regions. They therefore also thread through regions of weak disorder as shown from the broad distribution in Fig.~\ref{fig:SI2}(f). This can lead to the formation of bound states close to zero energy. Their subsequent overlapping lead to its delocalization of these states, eventually leading to a flat DOS for weakly disordered systems.

\end{document}